\documentclass[conference]{IEEEtran}
\IEEEoverridecommandlockouts
\usepackage{cite}
\usepackage{amsmath,amssymb,amsfonts}
\usepackage{float}
\usepackage{array}
\usepackage{hyperref}
\usepackage{algorithm}
\usepackage{multirow}
\usepackage{algcompatible}
\usepackage{algpseudocode}
\usepackage{graphicx}
\usepackage{textcomp}
\usepackage[table,xcdraw]{xcolor}
\usepackage{caption}
\usepackage{subcaption}
\def\BibTeX{{\rm B\kern-.05em{\sc i\kern-.025em b}\kern-.08em
    T\kern-.1667em\lower.7ex\hbox{E}\kern-.125emX}}
    
\begin{document}

\title{Distribution grid power flexibility aggregation at multiple interconnections between the high and extra high voltage grid levels\\
\thanks{The study performed under the research project "SiNED - System Services for secure electricity grids in times of advancing energy transition and digital transformation" acknowledges the support of the Lower Saxony Ministry of Science and Culture through the "Niedersächsisches Vorab" grant program (grant ZN3563) and of the Energy Research Centre of Lower Saxony.''}
}

\author{\IEEEauthorblockN{Neelotpal Majumdar, Marcel Sarstedt, Lutz Hofmann}
\IEEEauthorblockA{\textit{Institute of Electric Power Systems, Electric Power Engineering Section} \\
\textit{Leibniz Universität Hannover}\\
Hanover, Germany \\
majumdar@ifes.uni-hannover.de (https://orcid.org/0000-0002-4859-1247)}
}

% \and
% \and
% \and
% }

\maketitle

\begin{abstract}
The energy transition towards renewable based power provision requires improved monitoring and control of distributed energy resources (DERs), installed predominantly at the distribution grid level. Due to the gradual phase out of thermal generation, a shift of ancillary services provision like voltage control, congestion management and dynamic support from DERs is underway. Increased planning for procurement of ancillary services from underlying grid levels is required. Therefore, provision of flexible active and reactive power potentials from distribution system operators to transmission system operators at the vertical system interface is a subject of current research. At present, provision of active and reactive power flexibilities (PQ-flexibilities) across radial system interconnections are investigated, which involves a single transformer branch interconnection across two different voltage levels. Inclusion of multiple interconnections in a meshed grid structure increases the complexity as proximal interdependencies of interconnections to PQ-flexibilities require consideration. The objective of this paper is to address the flexibility aggregation across multiple vertical interconnections. Alternating current power transfer distribution factors (AC-PTDFs) are used to determine the power flow across the interconnections. Subsequent integration in a linear optimization environment controls the interconnection power flows (IPF) using a weighted objective function. Therefore, power flow regulation is enabled according to the requirements and specifications of both the underlying and overlaying grid level. The results show interdependent concentric flexibility regions or Feasible Operating Regions (FORs) in accordance with the manipulation of the weighted objective function.
\end{abstract}

\begin{IEEEkeywords}
ancillary services, active distribution grid, TSO-DSO cooperation, Feasible Operating Region, AC-PTDF, linear optimization
\end{IEEEkeywords}

\section{Introduction}
The phasing out of thermal power plants requires DERs to ensure a reliable and secure power supply. Therefore, the DSO requires extensive planning and operational management to provide ancillary services locally within the distribution grid and for the TSO at the overlaying transmission grid. Flexibility of active and reactive power supply (PQ-flexibility) can be ensured for example by the converter coupled DERs, predominantly installed at the distribution grid. To enable an exchange interface between the distribution and transmission grid level, flexibility mapping techniques from Active Distribution Networks (ADNs) are a subject of current studies. Researches have demonstrated stochastic scenario generation \cite{heleno2015estimation}\cite{gonzalez2018determination}\cite{ageeva2019analysis}\cite{riaz2019feasibility}, mathematical optimization \cite{rossi2017fast}\cite{silva2018estimating}\cite{contreras2018improved} and meta-heuristic optimization \cite{sarstedt2021survey} for determining the PQ-flexibility potential of ADNs. Such a flexibility mapped region is termed as the Feasible Operating Region (FOR) or the PQ-capability curves. Extensive surveys are presented in \cite{sarstedt2021survey}\cite{papazoglou2022review}, discussing the state of the art research in this field. Detailed comparison of varied methods are presented in \cite{contreras2018improved}\cite{sarstedt2021survey}\cite{majumdar2022linear}. Novel researches addressing uncertainties in power injections \cite{kalantar2019characterizing}\cite{wang2020stochastic}\cite{tan2020estimating}, state estimation in the context of FOR determination \cite{muller2021online}, are presented. The economic aspects considering monetarization of the FOR is also demonstrated in \cite{sarstedt2022monetarization}. However, the aspect of determining the FORs across multiple interconnection has not been addressed adequately.

Most of the researches have determined FORs aggregating flexibilities at the medium and high voltage (MV-HV) interconnection, considering radial distribution networks. However, there are multiple interconnections between the high voltage (HV) and extra high voltage (EHV) grid levels. Therefore, a corresponding determination of the FORs in a meshed grid with multiple interconnectors is of considerable importance. In \cite{kubispaper}, this issue is addressed, emphasizing the mutual dependency of flexibility provision at multiple interconnections. The method used requires cost optimal power flow calculations for a multitude of PQ-flexibility constellation scenarios at the interconnections, enabling pre-determined mutually interdependent flexibility mapping. Disadvantages however include simulation of a multitude of possible PQ-flexibility constellations. Furthermore, specified TSO-DSO coordinated exchanges at the interconnections, e.g., prioritized flexibility provision at a specific interconnection in relation to others is not addressed. To address this issues, the objective of this paper is to present a preliminary method for identifying interdependent FORs at multiple interconnections using AC-PTDFs. Application of AC-PTDFs are well established in power system studies \cite{lee1992distribution}\cite{singh1997improved}\cite{leveringhaus2014comparison}\cite{leveringhaus2015combined}. An AC-PTDF formulation is presented in this paper and correspondingly adapted in a multi-objective optimization method for the interdependent FOR determination. Regulation of the power flows across multiple interconnections is achieved through adjustment of the weighted objective function.
This paper is organized as follows. Section II presents the FOR determination concept across multiple interconnections. Section III discusses the involved mathematical formulations and description of the optimization algorithm in the context of the FOR determination. Section IV presents the simulation results for flexibility aggregation across a test grid with 3 vertical interconnections . Section V concludes the paper with a summary of the important points, followed by section VI, discussing the extended developments envisaged.

\section{The concept of FOR determination at multiple interconnections}

\begin{figure}
    \centering
    \includegraphics[width=9cm]{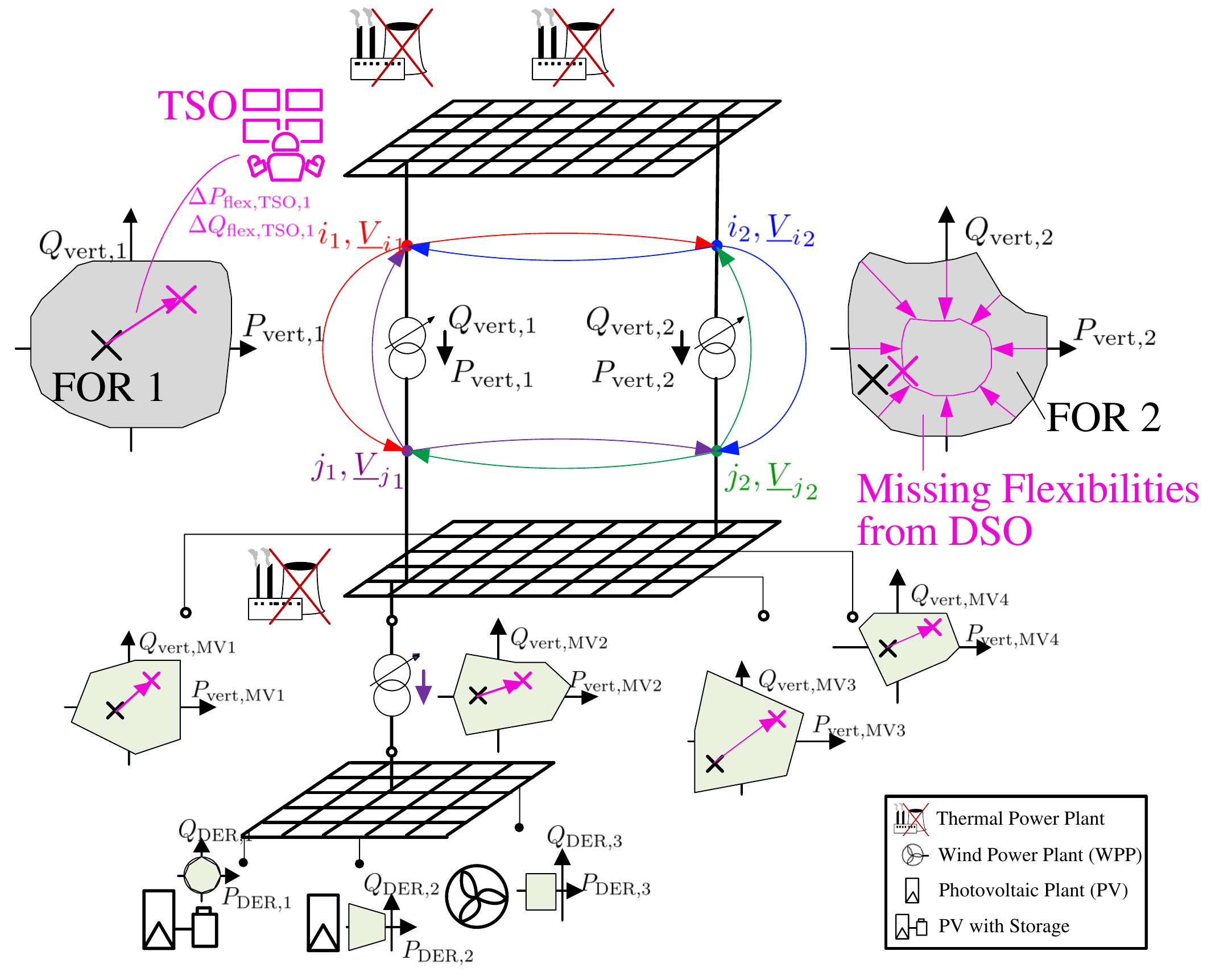}{}
    \caption{A schematic representation of flexibility aggregation (FOR determination) across multiple-grid levels}
    \label{fig:mulforconcept}
\end{figure}

Fig. \ref{fig:mulforconcept} illustrates the aggregation of flexibilities across multiple voltage levels with single and multiple interconnections. Typically high and extra high voltage levels (HV-EHV) are characterized by multiple interconnections. Distribution grid flexibilities from wind and solar plants primarily installed at the MV and LV grid levels usually assume a rectangular or triangular PQ-capability curve. Exemplary capability curves for wind and solar power plants, storage equipped solar power plants are demonstrated. Detailed recommendations for such variants of PQ-capabilities of DERs adhering to technical guidelines are discussed in \cite{majumdar2022linear}\cite{contreras2018improved}\cite{sarstedt2020simulation}. Such PQ-flexibilities from DERs are aggregated at the single MV-HV interconnection as demonstrated in Fig. \ref{fig:mulforconcept} ($f(P_\text{MV1},Q_\text{MV1})$, $f(P_\text{MV2},Q_\text{MV2})$, $f(P_\text{MV3},Q_\text{MV3})$, $f(P_\text{MV4},Q_\text{MV4})$). Across a single vertical interconnection, the process of aggregation is relatively simpler. A deviation of a DER operating point (active/reactive power), reflects equally at the MV-HV inter-connection, if losses are neglected. The process for multiple interconnections is more complex, as a deviation in the HV node PQ-flexibilities is divided across the interconnections. The power flow across the HV-EHV interconnections in the illustration are ($f(P_\text{vert,1},Q_\text{vert,1})$, $f(P_\text{vert,2},Q_\text{vert,2})$), for interconnections 1 and 2. Proximal PQ-flexibilities to interconnection 1 ($f(P_\text{MV1},Q_\text{MV1})$, $f(P_\text{MV2},Q_\text{MV2})$) have an increased influence as compared to the more remote flexibilities of ($f(P_\text{MV3},Q_\text{MV3})$, $f(P_\text{MV4},Q_\text{MV4})$). This is illustrated by the comparative decreased deviation in operating points of the proximal flexibilities, as compared to the remote flexibilities, to achieve a corresponding deviation at interconnection 1. Furthermore, the corresponding effect on the flexibility provision at interconnection 2 is demonstrated. An increase in flexibility potential from the distribution grid level at interconnection 1 has an effect of decreased flexibility provision at interconnection 2 (see pink area FOR 2 in Fig. \ref{fig:mulforconcept}). Such a decrement in the FOR presents a hypothetical scenario where the DSO is unable to comply with the TSO's specifications, thus jeopardizing system security in the transmission grid.

This paper addresses the proximal dependencies of the PQ-flexibilities on the multiple HV-EHV interconnections using AC-PTDFs to approximate the interconnection power flows. Subsequent usage of the IPFs for the interdependent flexibility aggregation (FOR determination) across the interconnections is achieved. An optimization algorithm is designed to adapt the power injections from the corresponding HV node PQ-flexibilities to determine the interdependent flexibility aggregation at multiple interconnections. In the simulations, power transfer through a specific interconnection is prioritised resulting in a corresponding decrease in flexibility provision through the other interconnections.

\section{Mathematical formulation of linearized sensitivities and the optimization algorithm}
\subsection{Linear sensitivities and power transfer distribution factors}

The linear programming approach presented in this paper requires linearization of the inherently non-linear and non-convex power flow equations. The linearized sensitivities are derived for approximating the grid power flow subject to deviations in power injection at the nodes. Subsequently, the power flow equations based on polar coordinate system are presented, with $v$ representing bus voltages, $y$ representing the network admittances. The bus voltage angles are represented by the symbol $\delta$ and $\theta$ represents the phase of the branch admittance.
\begin{equation}p_i = v_i^2y_{ii}\cos\theta_{ii}+\sum_{\substack{j=1 \\i\neq j}}^{n}v_iv_jy_{ij}\cos(\delta_i-\delta_j-\theta_{ij})\label{eq:1}\end{equation}
\begin{equation}q_i = -v_i^2y_{ii}\sin\theta_{ii}+\sum_{\substack{j=1 \\i\neq j}}^{n}v_iv_jy_{ij}\sin(\delta_i-\delta_j-\theta_{ij})\label{eq:2}\end{equation}
where $i,j \in n$ represent the bus or node indices. The power flow equations are approximated using the first order Taylor's polynomial approach. The corresponding inverse Jacobian sensitivities are obtained from the Newton Raphson power flow calculations\footnote[1]{Constant loads are considered in this research, neglecting voltage dependencies. Inclusion of voltage dependent loads requires an adaptaion of the shunt impedances in (\ref{eq:1}) and (\ref{eq:2}). However, in the context of the performed study, this enhancement is not deemed significant}.
\begin{equation}
\begin{bmatrix}
	\Delta \boldsymbol{\delta} \\
	\Delta \boldsymbol{v}
\end{bmatrix}=\boldsymbol{J}^{-1}\begin{bmatrix}
\Delta \boldsymbol{p} \\
\Delta \boldsymbol{q}
\end{bmatrix}\label{eq:3}\end{equation}
with \begin{equation*}
\renewcommand*{\arraystretch}{2.0}
\boldsymbol{J}^{-1}=\begin{bmatrix}
	\dfrac{\partial \boldsymbol{\delta} }{\partial \boldsymbol{p}} &\dfrac{\partial \boldsymbol{\delta} }{\partial \boldsymbol{q}} \\
	\dfrac{\partial \boldsymbol{v} }{\partial \boldsymbol{p}} & \dfrac{\partial \boldsymbol{v} }{\partial \boldsymbol{q}}
\end{bmatrix}\end{equation*}
Similarly, the branch power flow sensitivities or AC-PTDFs are derived in the context of the research. The corresponding branch power flow equations, derived at the terminal node $i$ for a branch connecting the nodes $i$ and $j$ are:
\begin{equation}p_{ij} = v_i^2y_{ii}\cos\theta_{ii}+v_iv_jy_{ij}\cos(\delta_i-\delta_j-\theta_{ij})\label{eq:4}\end{equation}
\begin{equation}q_{ij} = -v_i^2y_{ii}\sin\theta_{ii}+v_iv_jy_{ij}\sin(\delta_i-\delta_j-\theta_{ij})\label{eq:5}\end{equation}
The corresponding active power transfer distribution factors dependent on the branch terminal voltage angles $\delta_i$, $\delta_j$,and voltage magnitudes $v_i$ and $v_j$ are depicted as:
\begin{equation*}
\frac{\partial p_{ij}}{\partial \delta_i}, \frac{\partial p_{ij}}{\partial \delta_j}, \frac{\partial p_{ij}}{\partial v_i}, \frac{\partial p_{ij}}{\partial v_j}
\end{equation*}
Similarly, the reactive power distribution factors based on the branch terminal voltages and voltage angles are presented.
\begin{equation*}
\frac{\partial q_{ij}}{\partial \delta_i}, \frac{\partial q_{ij}}{\partial \delta_j}, \frac{\partial q_{ij}}{\partial v_i}, \frac{\partial q_{ij}}{\partial v_j}
\end{equation*}
The branch power sensitivities with respect to bus voltage magnitude and angles are referred to as $\boldsymbol{PQ_\text{T}DV}_{2b,2n}$. The dimension $2b$ refers to the active and reactive branch power flow sensitivities for the set of branches, and $2n$ refers to the set of $\delta_i$ and $v_i$ deviations. The subscript 'T' refers to branch terminal sensitivities. Although corresponding integration in the optimization environment is possible, the relation with respect to bus power injection is not intuitive. Therefore, a transformation of the sensitivities in terms of bus power injection is deemed important. The inverse Jacobian matrix is therefore applied from (\ref{eq:3}), as follows:
\begin{equation}
\boldsymbol{PQ_\text{T}PQ}_{2b,2n}=\boldsymbol{PQ_\text{T}DV}_{2b,2n} \text{x} \boldsymbol{J}^{-1}_{2n,2n}
\label{eq:6}\end{equation}
The formulation, therefore, yields active power distribution factors of the form:
\begin{equation*}
\frac{\partial p_{ij}}{\partial p_i}, \frac{\partial p_{ij}}{\partial p_j}, \frac{\partial p_{ij}}{\partial q_i}, \frac{\partial p_{ij}}{\partial q_j}
\end{equation*}
Similar derivation of the reactive power distribution factors based on branch terminal power injections are presented as:
\begin{equation*}
\frac{\partial q_{ij}}{\partial p_i}, \frac{\partial q_{ij}}{\partial p_j}, \frac{\partial q_{ij}}{\partial q_i}, \frac{\partial q_{ij}}{\partial q_j}
\end{equation*}
The power injection based active and reactive power distribution factors are correspondingly implemented into the optimization environment. These AC-PTDFs are segregated based on the active and reactive power injections as $\boldsymbol{PQ_\text{T}P}_{2b,n}$ and $\boldsymbol{PQ_\text{T}Q}_{2b,n}$ respectively.

Additional current sensitivities are adapted from established researches \cite{leveringhaus2014comparison}\cite{leveringhaus2015combined}, to incorporate adherence of branch currents to the thermal current limit constraints. 
\begin{equation}
\Delta \boldsymbol{i} = \boldsymbol{ID}_\text{TT}\Delta\boldsymbol{\delta}_\text{T} + \boldsymbol{IU}_\text{TT}\Delta\boldsymbol{v}_\text{T}
\label{eq:7}\end{equation} 
% with \begin{equation*}
% \boldsymbol{ID}_\text{TT}=\dfrac{\partial \Delta \boldsymbol{i}_\text{T}}{\partial \boldsymbol{\delta}_\text{T}}; \boldsymbol{IU}_\text{TT}=\dfrac{\partial \Delta  \boldsymbol{i}_\text{T}}{\partial \boldsymbol{v}_\text{T}}
% \end{equation*}
The subscript TT refers to the branch terminal indices represented by node indices $i,j$. The terminal sensitivity matrices are adapted for integration in the optimization environment, as the variables used are bus voltages and phase angles. Therefore, sensitivity matrices of terminal currents to bus variables ($\boldsymbol{ID}_{\text{TB}}$,$\boldsymbol{IU}_{\text{TB}}$) is obtained by multiplication with the nodal incidence matrix, similar to \cite{majumdar2022linear}.

\subsection{Formulation of the optimization algorithm adapted for multiple interconnection FOR determination}

The multi-objective optimization is formulated as a linear programming algorithm. The optimal power flow is simulated for monotonically increasing active power deviation samples at the HV buses. The corresponding maximum positive and negative reactive power transfer through the vertical interconnections is determined for each sample. Therefore, the dependency of reactive power transfer for increasing active power deviations is captured.

Corresponding optimization slack variables for the AC-PTDFs are introduced, analogous to \cite{majumdar2022linear} for prioritization of the power flow through specified interconnections. The corresponding weights of the slack variables are adjusted in the objective function. The slack variable formulations is presented in the form of equality conditions:
\begin{equation}
\begin{split}
\boldsymbol{PQ_\text{T}P}_{2ic,n}\Delta \boldsymbol{p}+\boldsymbol{PQ_\text{T}Q}_{2ic,n}\Delta \boldsymbol{q}-\Delta \boldsymbol{x}_{\text{slack},2ic} = 0
\end{split}
\label{eq:8}\end{equation} with the linearized deviations in vertical branch power flows expressed as: \begin{equation*}
    \begin{split}
\boldsymbol{PQ_\text{T}P}_{2ic,n}\Delta \boldsymbol{p}+\boldsymbol{PQ_\text{T}Q}_{2ic,n}\Delta \boldsymbol{q}=\begin{bmatrix}
    \Delta \boldsymbol{p_\text{vert}}\\
    \Delta \boldsymbol{q_\text{vert}}
\end{bmatrix}
\end{split}
\end{equation*}
The subscript $ic$ refers to the set of inter-connections, for which the active and reactive power branch flow sensitivities are determined. The subscript 'vert' refers to the vertical active and reactive power branch flows. 

The optimization formulation is presented as follows:
\begin{equation}
\renewcommand*{\arraystretch}{1.5}
\begin{gathered}
\begin{split} \text{min} \quad & \boldsymbol{c}^\text{T}\boldsymbol{x} \mid \boldsymbol{x} = [\Delta \boldsymbol{p}^\text{T}, \Delta \boldsymbol{q}^\text{T}, \Delta \boldsymbol{\delta}^\text{T}, \Delta \boldsymbol{v}^\text{T}, \boldsymbol{x}_\text{slack}^\text{T}]^\text{T}_{m,1};\\ & \boldsymbol{c}=[c_k]^\text{T}_{m,1}, k = [1,m] \cap \mathbb{Z}; m = 4n+2ic\end{split}
\\ \boldsymbol{A}_{\text{ineq}}\boldsymbol{x}\leq\boldsymbol{b}_{\text{ineq}}; \boldsymbol{A}_{\text{eq}}\boldsymbol{x}=\boldsymbol{b}_{\text{eq}}\\\begin{split}\text{s.t} \quad & \Delta \boldsymbol{p}_{\text{min}} \leq \Delta \boldsymbol{p} \leq \Delta \boldsymbol{p}_{\text{max}} \\ & \Delta \boldsymbol{q}_{\text{min}} \leq \Delta \boldsymbol{q} \leq \Delta \boldsymbol{q}_{\text{max}}\\\end{split}
\end{gathered}\label{eq:opt}\end{equation}
$\boldsymbol{c}$ refers to the vector of objective function costs $c_k$ (or weights). The subscripts 'ineq' and 'eq' refers to the inequality and equality conditions respectively. The corresponding matrices are split into upper bound (subscript 'ub') and lower bound ('lb') conditions. The upper bound and lower bound conditions specify adherence to the maximum and minimum grid constraints respectively.\\\begin{equation*}
\renewcommand*{\arraystretch}{1.5}
\begin{gathered}
\boldsymbol{A}_{\text{ineq}}=\begin{bmatrix}
\boldsymbol{A}_{\text{ineq,ub}} \\ \boldsymbol{A}_{\text{ineq,lb}} 
\end{bmatrix};  \boldsymbol{b}_{\text{ineq}}=\begin{bmatrix}
\boldsymbol{b}_{\text{ineq,ub}} \\ \boldsymbol{b}_{\text{ineq,lb}} 
\end{bmatrix} \\  \boldsymbol{A}_{\text{eq}}=\begin{bmatrix}
\boldsymbol{A}_{\text{eq,ub}} \\ \boldsymbol{A}_{\text{eq,lb}} 
\end{bmatrix};  \boldsymbol{b}_{\text{eq}}=\begin{bmatrix}
\boldsymbol{b}_{\text{eq,ub}} \\ \boldsymbol{b}_{\text{eq,lb}} 
\end{bmatrix} \\  \boldsymbol{A}_{\text{ineq,lb}}=  -\boldsymbol{A}_{\text{ineq,ub}}; \boldsymbol{A}_{\text{eq,lb}}=  -\boldsymbol{A}_{\text{eq,ub}}\\  \boldsymbol{b}_{\text{ineq,lb}}=  -\boldsymbol{b}_{\text{ineq,ub}}; \boldsymbol{b}_{\text{eq,lb}}=  -\boldsymbol{b}_{\text{eq,ub}}\end{gathered}\end{equation*}\\ The matrices are accordingly expressed as:
\begin{equation*}
\renewcommand*{\arraystretch}{1.0}
\begin{gathered}
\boldsymbol{A_\text{{ineq,ub}}}= \begin{bmatrix} \dfrac{\partial \boldsymbol{\delta}}{\partial \boldsymbol{p}} & \dfrac{\partial \boldsymbol{\delta}}{\partial \boldsymbol{q}} & \boldsymbol{0}_{n,n} & \boldsymbol{0}_{n,n} & \boldsymbol{0}_{n,2ic} \\ \dfrac{\partial \boldsymbol{v}}{\partial \boldsymbol{p}} & \dfrac{\partial \boldsymbol{v}}{\partial \boldsymbol{q}} & \boldsymbol{0}_{n,n} &\boldsymbol{0}_{n,n} & \boldsymbol{0}_{n,2ic}\\ \boldsymbol{0}_{2b,n} & \boldsymbol{0}_{2b,n} & \boldsymbol{ID}_{\text{TB}} & \boldsymbol{IU}_{\text{TB}} & \boldsymbol{0}_{2b,2ic}\\
\boldsymbol{PQ_\text{T}P} & \boldsymbol{PQ_\text{T}Q} &\boldsymbol{0}_{2ic,n} &\boldsymbol{0}_{2ic,n} & \text{d}(\boldsymbol{-1}_{2ic,2ic}) \end{bmatrix}\\ \boldsymbol{A_\text{{eq,ub}}}= \begin{bmatrix}
 \dfrac{\partial \boldsymbol{\delta}}{\partial \boldsymbol{p}} & \dfrac{\partial \boldsymbol{\delta}}{\partial \boldsymbol{q}} & \text{d}(\boldsymbol{-1}_{n,n}) & \boldsymbol{0}_{n,n} & \boldsymbol{0}_{n,2ic} \\ \dfrac{\partial \boldsymbol{v}}{\partial \boldsymbol{p}} & \dfrac{\partial \boldsymbol{v}}{\partial \boldsymbol{q}} & \boldsymbol{0}_{n,n} &\text{d}(\boldsymbol{-1}_{n,n}) & \boldsymbol{0}_{n,2ic}\\ \boldsymbol{0}_{2b,n} & \boldsymbol{0}_{2b,n} & \boldsymbol{0}_{2b,n} & \boldsymbol{0}_{2b,n} & \boldsymbol{0}_{2b,2ic}\\ \boldsymbol{0}_{2ic,n} & \boldsymbol{0}_{2ic,n} & \boldsymbol{0}_{2ic,n} & \boldsymbol{0}_{2ic,n} & \boldsymbol{0}_{2ic,2ic}\end{bmatrix}\\ \boldsymbol{b}_{\text{ineq,ub}}=\begin{bmatrix}
\boldsymbol{\delta}_\text{max} - \boldsymbol{\delta}_0\\ 
\boldsymbol{v}_\text{max} - \boldsymbol{v}_0\\ 
\dfrac{\boldsymbol{i}_\text{max} - \boldsymbol{i}_0}{\boldsymbol{i}_0}\\ \boldsymbol{0}_{2ic,1}\\ 
\end{bmatrix}; \boldsymbol{b}_{\text{eq,ub}}=\begin{bmatrix}
\boldsymbol{0}_{n,1}\\ \boldsymbol{0}_{n,1}\\ \boldsymbol{0}_{n,1}\\ \boldsymbol{0}_{2ic,1} 
\end{bmatrix} \end{gathered}\end{equation*}\\
The dimension $m=4n+2ic$ refers to the four grid variables $\boldsymbol{\Delta p,\Delta q,\Delta \delta,\Delta v}$ and the $2ic$ slack variables for the inter-connector active and reactive power sensitivities. The dimensions for the matrix entries are specified for improved understanding of the formulation. The operator $\text{d}(\boldsymbol{-1})$ represents a diagonal of $-1$ with the specified matrix dimensions. The subscripts 'max', 'min' and '0' refer to the maximum, minimum and current operating point values for the constraints. It is to be noted that the angle constraints are specified for completeness of the formulation.

\subsection{FOR determination across multiple interconnections}

\begin{figure}[ht]
	\includegraphics[width=9cm]{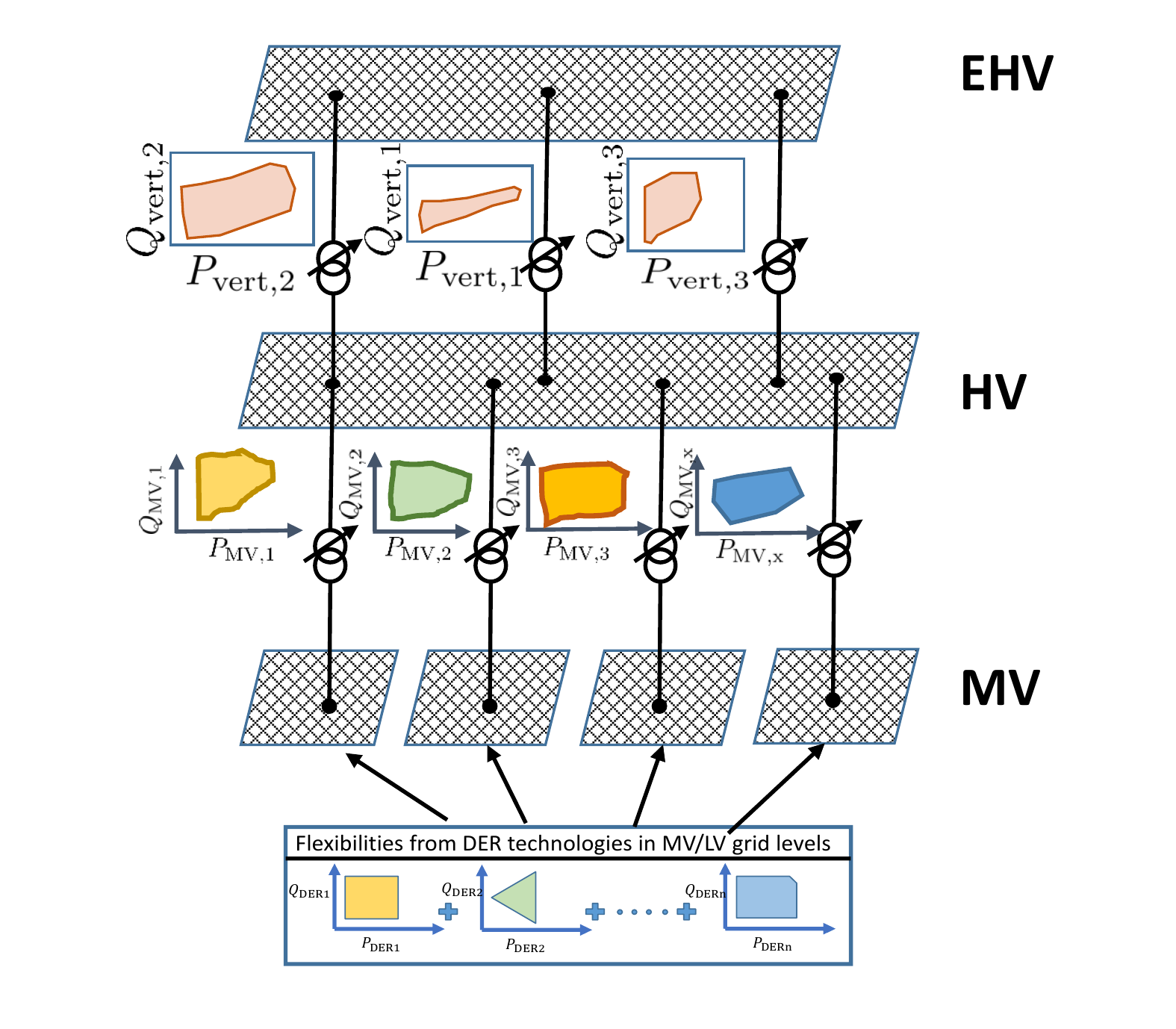}
	\caption{A schematic representation of a multi-voltage level grid flexibility aggregation at vertical interconnections}
	\label{fig:mulfor}
\end{figure}

The process of distribution grid flexibility aggregation is presented in Fig. \ref{fig:mulfor}. In the scope of this study, typical rectangular PQ-capability curves for the wind and solar power plants at MV and LV grid levels are assumed (see Section II). Aggregation of the corresponding flexibilities at each MV-HV interconnection is performed, according to the optimization formulation presented in \cite{majumdar2022linear}. Corresponding non-linear and non-convex Feasible Operating Regions (FOR) subject to local grid constraints are therefore obtained at the MV-HV interconnections, as displayed in Fig. \ref{fig:mulfor}. Subsequently, the flexibilities from the MV-HV interconnections acquired at the HV nodes require to be aggregated at the HV-EHV interconnections. The corresponding AC-PTDFs are determined for the interconnectors as previously mentioned. The interconnection PQ-flexibility potentials are regulated by adjusting the weights of the AC-PTDF slack variables in the objective function, according to the required prioritization.
The optimization is performed in steps or discrete samples. The active power potential for each HV-node flexibility e.g., for bus index 1, $\Delta p_{d=1}$ is known. Index $d \in n$ represents the HV-nodes with allocated DER power flexibilities. A corresponding discretization is performed  for $k_\text{max}$ number of samples, $\Delta p_{d=1}/k_\text{max}$. This step is undertaken for all flexibilities at the HV-nodes. The optimization is performed for $k = [1,k_\text{max}]$, determining the maximum and minimum reactive power potentials at the HV-nodes, for sequentially sampled active power deviations.

\begin{figure}
	\centering
	\begin{tabular}{cc}
		\subfloat[Exemplary MV-HV interconnection flexibility at HV node x]{\includegraphics[width=0.23\textwidth]{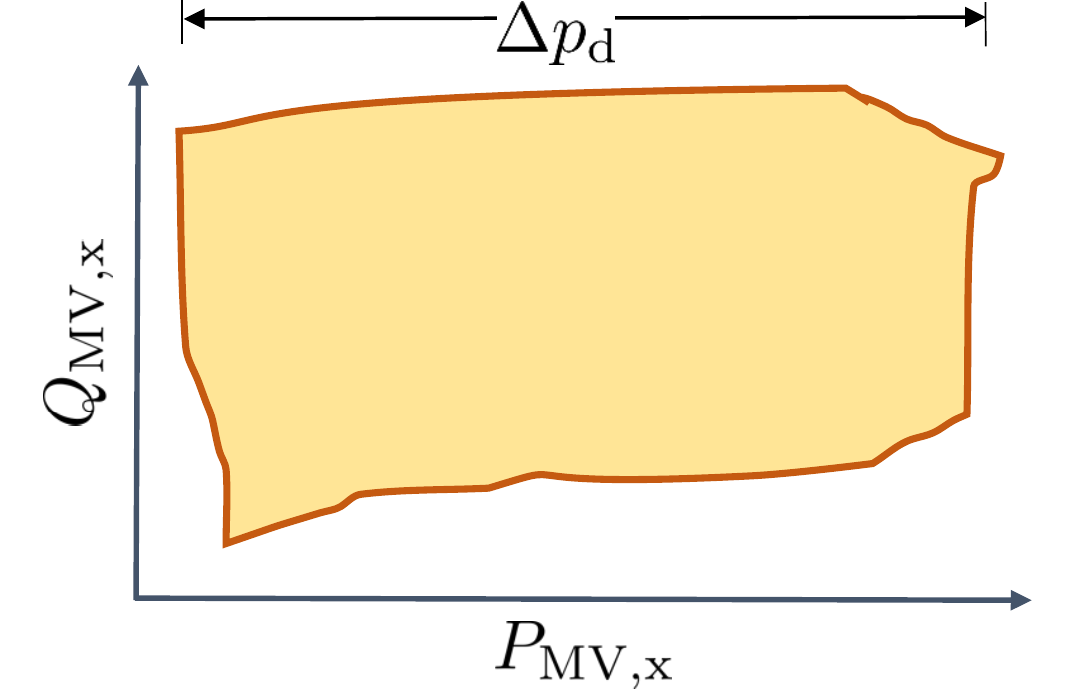}}
		&
		\subfloat[Discretization of the exemplary flexibility potential]{\includegraphics[width=0.23\textwidth]{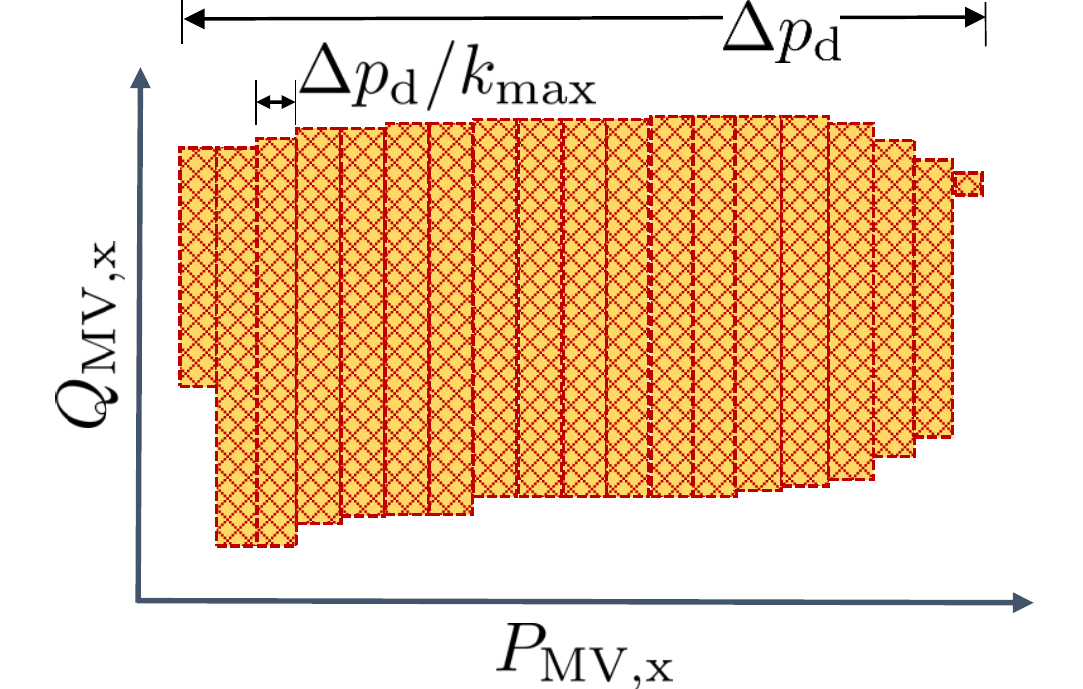}}
	\end{tabular}
	\caption{Exemplary HV-node flexibility potential and corresponding discretization}
	\label{fig:exemflex}
\end{figure}

The corresponding discretization of an exemplary HV node flexibility is illustrated in Fig. \ref{fig:exemflex}. Observations reveal a fixed amount of maximum and minimum reactive power potential for each step-wise increment of active power deviation along the x-axis. The non-linear shape in Fig. \ref{fig:exemflex} (a) can indeed be approximated using a piece-wise linearization approach. However, to address the focus of this paper a simplified discretization as discussed, is presented to demonstrate the preliminary results.
In order to determine the maximum/minimum reactive power potential at the interconnections, a threshold specification of reactive power injection from HV node flexibilities is required. A maximum/minimum injection of Q-flexibility from the nodes, determines a corresponding maximum/minimum Q-potential transfer at the HV-EHV vertical interconnection. This reactive power transfers across the vertical interconnections are constant values as the flexibility injection is constant (maximum/minimum). Therefore, to analyse the application of AC-PTDFs for influencing the power flows a corresponding constraint  or threshold for reactive power flexibility utilization is required. The optimization formulation from (\ref{eq:opt}) is compounded with (\ref{eq:Qthresh}), where $Q_\mathrm{Thresh}$ is a pre-specified value limiting the total sampled reactive power injection $\sum_{d=1}^{n} \Delta q_{d,\text{samp}}$ from the HV node flexibilities. Different values of $Q_\mathrm{Thresh}$ can be analysed as case studies to determine different combinations of vertical reactive IPFs.
\begin{equation}
    \sum_{d=1}^{n} \Delta q_{d,\text{samp}}-Q_\text{Thresh}=0
    \label{eq:Qthresh}
\end{equation}
Corresponding constraints can be placed on AC-PTDF IPF slack variables, to allow a fixed or maximum/minimum amount of power transfer. For sampled active power flows, the weights in the objective function are negative, for maximum positive incrementally sampled active power deviation.
\begin{equation}
\Delta p_{d,\text{samp}}=\frac{\Delta p_d}{k_\text{max}}.
\label{eq:pdsample}
\end{equation}
This enables incremental step-wise down-regulation of active power, subject however, to the local grid constraints for bus voltages and line thermal current limits. The active power flows can be adjusted according to specifications by placing constraints on the corresponding slack variables. In practice, specifications can be recommended by the transmission system operator (TSO) or the corresponding overlaying grid operator to the underlying distribution grid operator (DSO). This method therefore, enables an effective TSO-DSO based coordination between multiple-grid levels for operational management of ancillary services. In the undertaken research the focus is on the reactive power potential adjustment, for simplified changes in active power flows. For an intricate analysis, piece-wise linearization of the non-linear flexibility maps are recommended. The algorithm for the FOR determination across multiple interconnections is presented in Algorithm \ref{alg:cap}.
\begin{algorithm}
\begin{algorithmic}[1]
\caption{Determination of HV-EHV interconnection FORs}\label{alg:cap}
\State Aggregate the MV-HV interconnection flexibilities at HV nodes with the formulation presented in \cite{majumdar2022linear}
\State Discretize the local HV node DER flexibilities and the flexibilities aggregated for each HV node
\While{$k \leq k_\text{max}$}
\State Determine $max$ +ve $Q_\text{vert}$ according to specified objective priority for incremental $\Delta p_{d,\mathrm{samp}}$ power deviation
\State Adjust objective weights of the IPF slack variables
\State $Q_\text{Thresh} \gets$ $+ve$ specified value;
\State Solve optimization (\ref{eq:opt}) compounded with (\ref{eq:Qthresh}), (\ref{eq:pdsample})
\State Update operating points of the flexibilities $p_d$, $q_d$
\EndWhile
\While{$k \leq k_\text{max}$}
\State Determine $max$ -ve $Q_\text{vert}$ according to specified objective priority for incremental $\Delta p_{d,\mathrm{samp}}$ power deviation
\State Adjust objective weights of the IPF slack variables
\State $Q_\text{Thresh} \gets$ $-ve$ specified value;
\State Solve optimization (\ref{eq:opt}) compounded with (\ref{eq:Qthresh}), (\ref{eq:pdsample})
\State Update operating points of the flexibilities $p_d$, $q_d$
\EndWhile
\end{algorithmic}
\end{algorithm}

\subsection{Validation of the AC-PTDFs with Newton-Raphson load flow method}

\begin{figure}
	\centering
	\begin{tabular}{cc}
		\subfloat[Deviation of approximated $q_{ij}$ for IPFs with incremental $p_i$ deviations]{\includegraphics[width=0.2\textwidth]{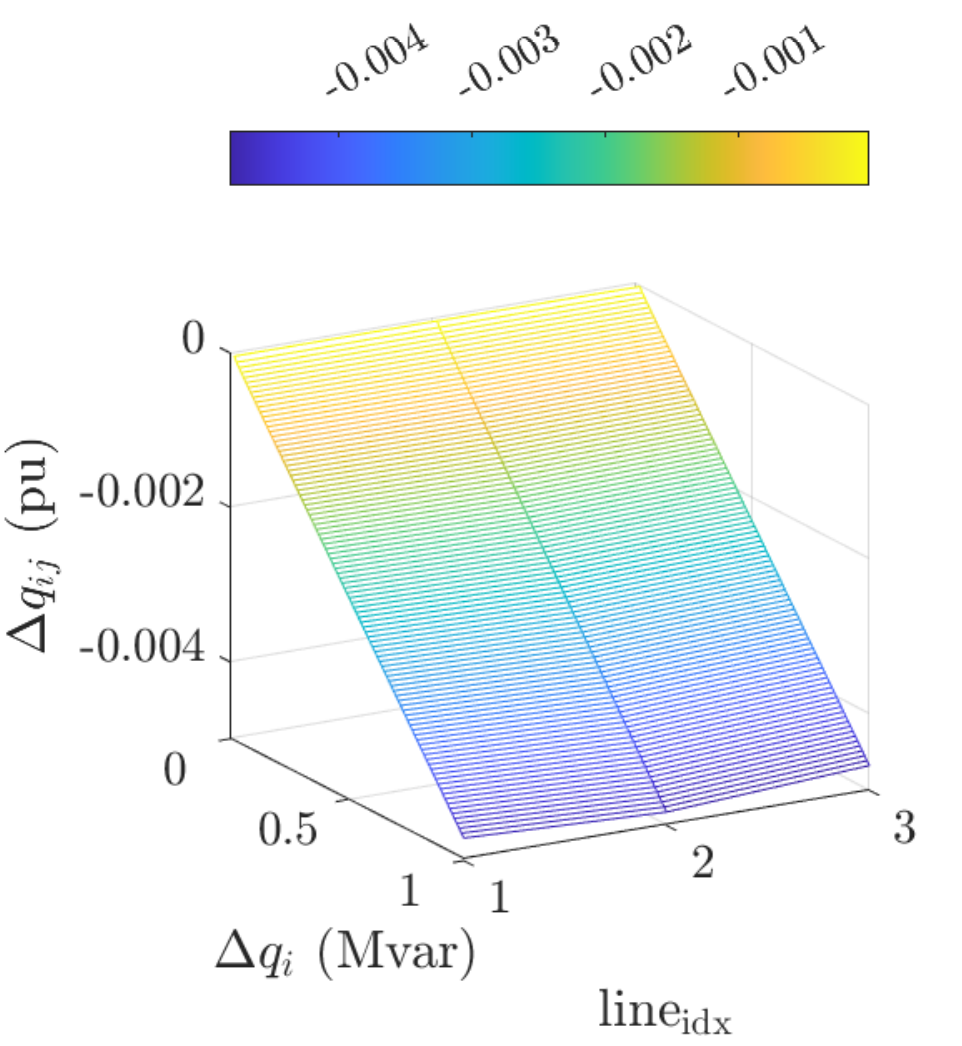}}
		&
		\subfloat[Deviation of approximated $q_{ij}$ for IPFs with incremental $q_i$ deviations]{\includegraphics[width=0.2\textwidth]{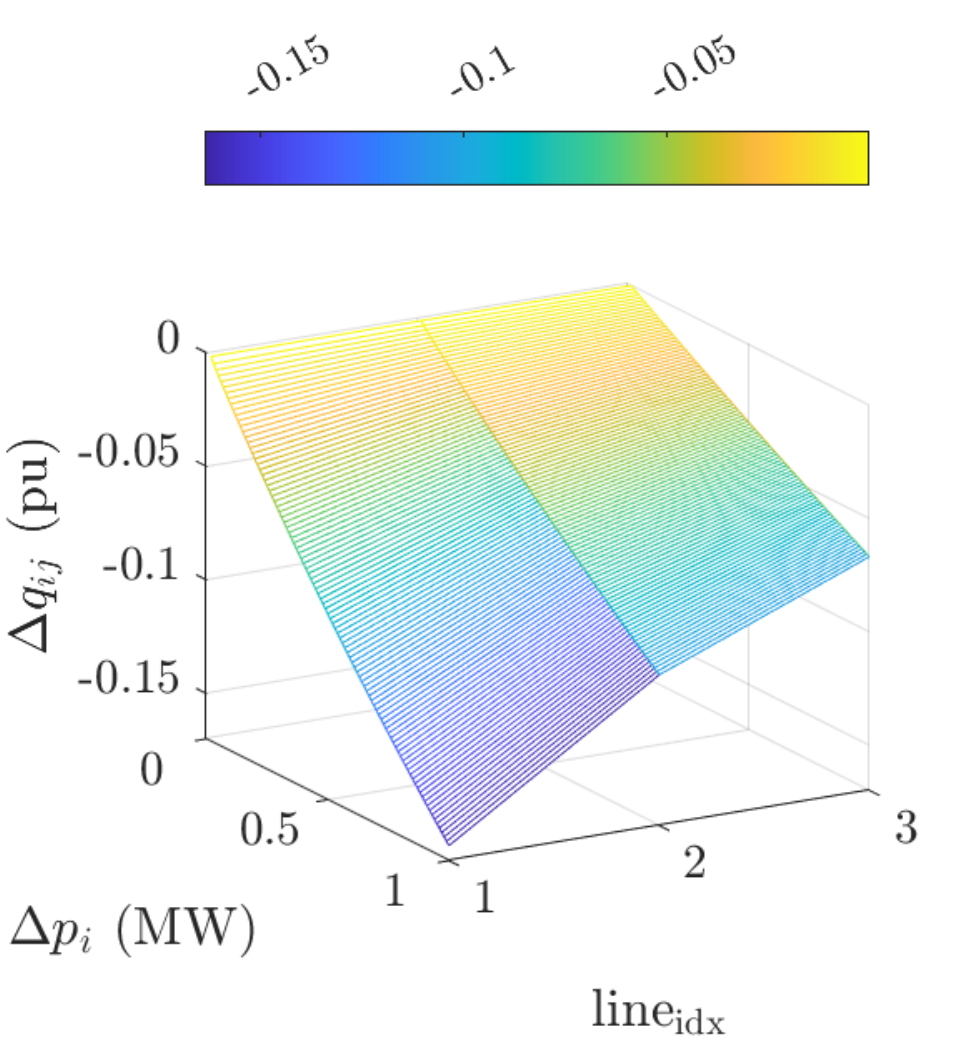}}
	\end{tabular}
	\caption{Inter-connection branch flow AC-PTDFs as a function of active and reactive bus power injections}
	\label{fig:acptdf}
\end{figure}

The application of AC-PTDFs for approximation of power flows is demonstrated in a multitude of research. A particularly insightful analysis is presented in \cite{coffrin2014linear}, demonstrating the influence of angle and voltage deviation on active and reactive power flows. A simulative illustration of AC-PTDFs in the scope of the undertaken research is presented in Fig. \ref{fig:acptdf}. Comparisons are based on the reactive power flow deviations $\Delta q_{ij}$ obtained as a normalized difference of AC-PTDF approximations and value obtained from power flow calculations. 
\begin{equation}
    \Delta q_{ij}=\frac{q_{ij,\text{AC-PTDF}}-q_{ij,\text{power flow}}}{q_{ij,\text{power flow}}}
\end{equation}
Observations reveal that interconnection reactive power flow deviations $\Delta q_{ij}$, with increased $q_i$ injections range within $-0.5 \%$. In comparison, the deviation with regards to incremental $p_i$ injections range within $-17 \%$. For comparability, the deviations in $p_i$ and $q_i$ are equal in magnitude. In the scope of this research, the FOR determination involves sampled active power increments. The $p_i$ deviations are reduced and therefore, resulting errors in power flow determination are deemed insignificant.

\subsection{Multi-level HV-EHV grid model for flexibility provision studies}

\begin{figure}[ht]
	\includegraphics[width=9cm]{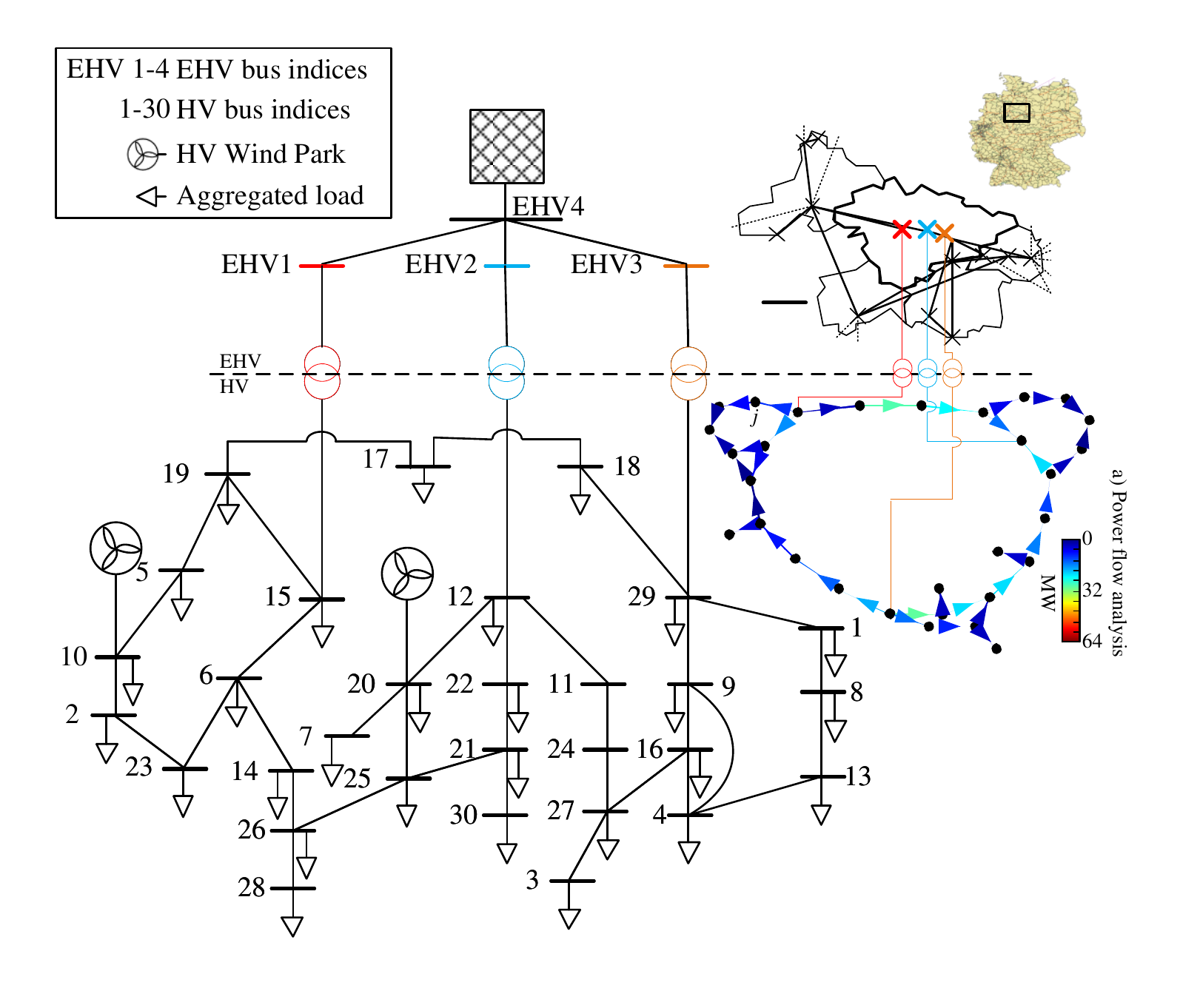}
	\caption{The examined EHV-HV grid level and topological description in a hierarchical grid coordination context}
	\label{fig:grid}
\end{figure}

The flexibility aggregation analyses across multiple vertical interconnections is performed on the HV-EHV grid model in Fig. \ref{fig:grid}. The grid is a part of a multi-voltage level grid framework consisting of multiple MV grid levels, HV grid and EHV grid level, presented in \cite{sarstedt2019modelling}. The power flow and grid dataset is established in \cite{ifes-eevdataset}, which is adapted according to the proposed bottom-up flexibility aggregation. The aggregated flexibility potentials at corresponding MV-HV grid vertical interconnections is obtained from the underlying MV grid levels, as previously mentioned in sub-section II. C. Subsequent analyses is performed considering the MV-HV grid flexibilities as corresponding flexibility potentials at the HV nodes. The HV grid level is at 110 kV. The original grid consists of 30 HV nodes and 3 EHV nodes (1-3) as observed from Fig. \ref{fig:grid}. A topological relation with a representative power flow scenario is presented with reference to the location in Germany. The EHV 4 bus is added as a slack bus for power flow calculations, and to derive AC-PTDFs across the 3 vertical interconnections (interconnection 1: EHV 1- HV 15; interconnection 2: EHV 2- HV 12; interconnection 3: EHV 3- HV 29). 

\section{Simulation Results}

The simulation is segregated into 3 scenarios and 2 case studies. The case studies are performed considering a maximum reactive power flexibility utilization ($\pm Q_\mathrm{Thresh}=60$ Mvar and $\pm Q_\mathrm{Thresh}=40$ Mvar). For each case study, 3 scenarios are defined to determine particular combinations of FORs across the vertical interconnections. The scenarios described are classified according to priority of increased flexibility provision at a selected interconnection in comparison to the others. The prioritization is controlled by adjusting the weights of the slack variables associated with the respective AC-PTDFs, as previously mentioned.

\begin{itemize}
    \item Scenario 1: Flexibility provision is maximised at vertical interconnection 1.
    \item Scenario 2: Flexibility provision is maximised at vertical interconnection 2.
    \item Scenario 3: Flexibility provision is maximised at vertical interconnection 3.
\end{itemize}

\begin{figure}[ht]
	\includegraphics[width=9cm]{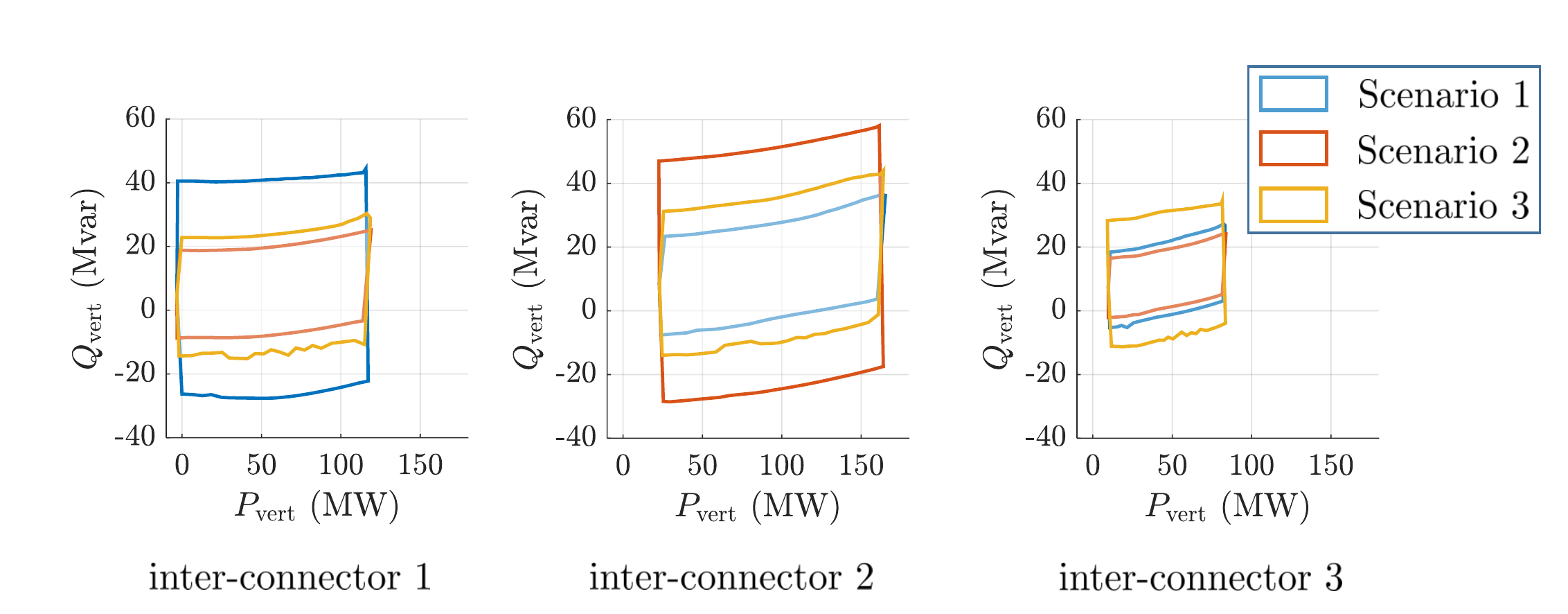}
	\caption{Flexibility provision potential at multiple inter-connections for $Q_\text{Thresh}=$ 60 Mvar}
	\label{fig:flexcase1}
\end{figure}

\begin{figure}[ht]
	\includegraphics[width=9cm]{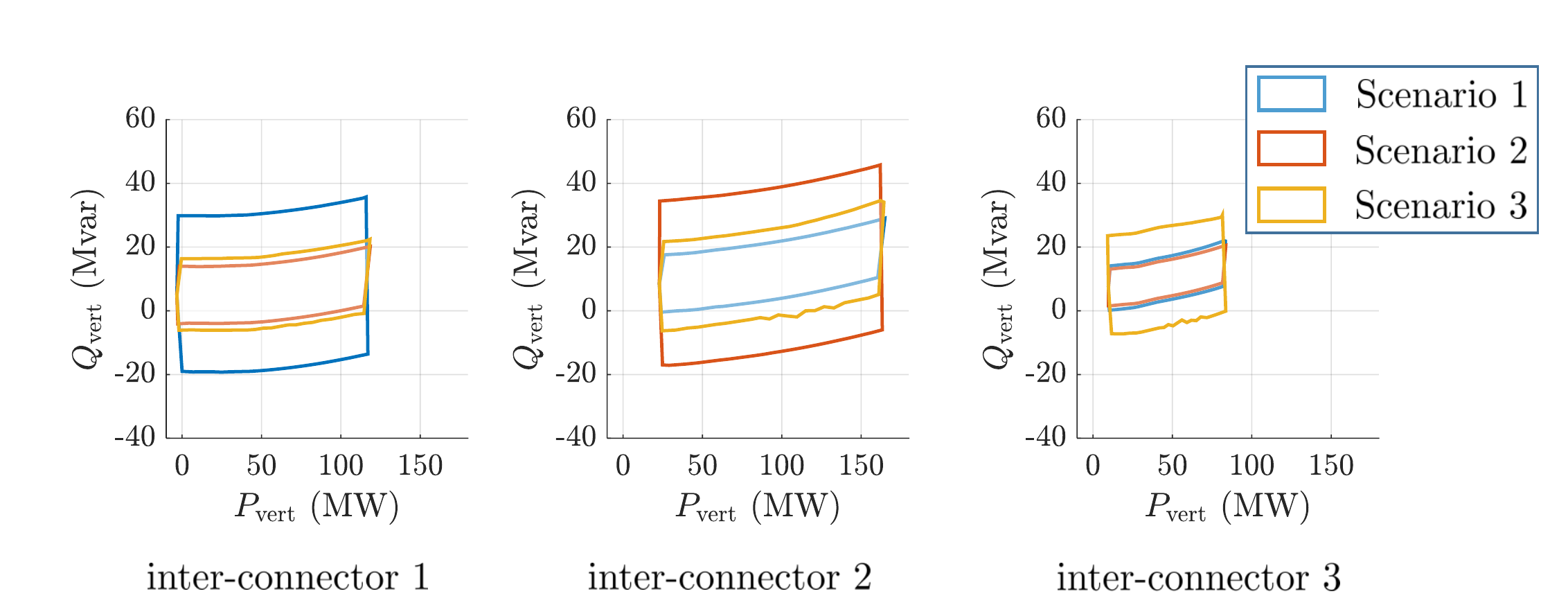}
	\caption{Flexibility provision potential at multiple inter-connections for $Q_\text{Thresh}=$ 40 Mvar}
	\label{fig:flexcase2}
\end{figure}

Fig. \ref{fig:flexcase1} illustrates the examined scenarios for case study 1 ($Q_\mathrm{Thresh}=60$ Mvar). Investigations reveal that the flexibility provision is maximised for the specified interconnections in accordance with the scenarios presented. Scenario 1 maximizes reactive power transfer across interconnecton 1, resulting in a reduced transfer at interconnections 2 and 3 (blue perimeter). Similarly, Scenario 2 maximizes power transfer across interconnection 2 while reducing flexibility potential at others (indicated by the red perimeter). Scenario 3 enhances PQ-flexibility provision at interconnection 3 (as indicated by the yellow perimeter). The corresponding results for case study 2 ($Q_\mathrm{Thresh}=40$ Mvar) is presented in Fig. \ref{fig:flexcase2}. A similar specification of flexibility demand from the transmission grid operator may differ at different regions (connected to different interconnections). Prioritization of flexibility at specific regions can arise due to localized voltage or congestion problems. Therefore, such a TSO-DSO coordination scheme is deemed useful to address the region-specific variation in flexibility demand.

\section{Conclusion}

The undertaken research deals with aggregation of flexibility potentials from the underlying HV grid at multiple vertical interconnections to the overlying EHV grid level. Flexibility aggregation across multiple interconnections is complex as proximal influence of flexibilities to neighbouring interconnections require consideration. Transfer of power from flexibilities to the vertical interconnections decreases with increase in the distance, due to increased power loss. To capture this proximal dependency, AC-PTDFs are developed to approximate the interconnection power flows as an effect of HV node bus power injections. A multi-objective optimization method with integrated AC-PTDFs is proposed to utilize the interconnection power flow interdependencies for prioritized power transfer across specific interconnections. Results reveal the interdependencies of the vertically aggregated Feasible Operating Regions (FORs) and prove the method's efficacy in maximising the flexibility potential across specified interconnections.

\section{Discussion on Extended Developments}

The presented approach can be interpreted as a first step for aggregating interdependent flexibility potential across multiple interconnections. Increasing the level of detail (e.g. enhancing the granularity of $Q_\mathrm{Thresh}$) results combinatorially in a multitude of aggregation runs. Simulation of varied possible vertical power flow scenarios can be visualized as concentric FORs to represent the combinations.
A flexibility provision from the distribution grid level corresponds to change of the grid state in the transmission grid level. 
% Even in the case of a single vertical system interface, the changed grid state in the transmission grid in turn influences the grid state in the distribution grid level, which shifts the boundaries of the FOR. To take this interaction into account, a PQ(V)-FOR is proposed in \cite{schwerdfeger2017vertikaler}\cite{sarstedt2020simulation}. This represents the voltage dependency of the FOR and can be used by the TSO to independently check the guarantee of flexibility provision from the distribution grid level. 
In addition to this vertical dependency of the grid state variables ($v, \delta$) between transmission and distribution network level, the presented approach neglects the horizontal dependency within the transmission grid level, since the vertical system interfaces are connected with a common slack bus. In general, the consideration of these dependencies is also possible via AC-PTDF, which can be provided to the DSO by the TSO considering the grid data protection aspects. The FOR determination at multiple vertical system interfaces results in a large number of additional aggregation runs, which can be interpreted as additional dimensions of the FOR. Thus, a future challenge is both to determine such a Hyperspace-FOR in a performant way and to integrate it into the TSO's operational planning. The importance of this research field continues to increase considering additional information within the FOR. Examples are the consideration of uncertainties in the load and generation forecast, technical characteristics of the FPU (e.g. start-up times, operating time) or flexibility prices with regard to a flexibility market based TSO-DSO cooperation. 
% \begin{figure}[ht]
% 	\includegraphics[width=9cm]{MarcelFORscheme.pdf}
% 	\caption{A schematic representation of FOR inter-dependency}
% 	\label{fig:Marcelfor}
% \end{figure}

\bibliographystyle{ieeetr}
\bibliography{citation}

\end{document}